\documentclass[prd,aps,eqsecnum,floatfix,nofootinbib,preprint,tightenlines,showpacs]{revtex4}
\usepackage{latexsym}
\usepackage{hyperref}
\def\a{\alpha}
\def\b{\beta}

\def\d{\delta}
\def\g{\gamma}

\def\m{\mu}
\def\n{\nu}

\def\r{\rho}                    %       \varrho
\def\s{\sigma}                  %       \varsigma

\def\x{\xi}

\def\D{\Delta}

\def\S{\Sigma}

\def\cd{{\cal D}}

\def\cf{{\cal F}}
\def\cg{{\cal G}}
\def\ch{{\cal H}}   % overridden by cosh !!

\def\cl{{\cal L}}

\def\cw{{\cal W}}

\def\cbo{{\,\raise-.15ex\Sc [\,}}                       % curly "
\def\ddt#1{{\buildrel {\hbox{\LARGE .\kern-2pt.}} \over {#1}}}% double dot-over
\def\beqn#1{ \renewcommand{\theequation}{#1}
             \begin{eqnarray} }
\def\eeqn{ \renewcommand{\theequation}{\arabic{equation}}
           \end{eqnarray} }

\def\beqr#1{ \setcounter{equation}{#1}
             \begin{eqnarray} }

\def\eeqr{\end{eqnarray}}

\def\beqrabc#1{ \setcounter{equation}{0}
                \renewcommand{\theequation}{#1\alph{equation}}
                \begin{eqnarray} }
\def\beqrn#1#2{ \setcounter{equation}{#2}
                \renewcommand{\theequation}{#1.\arabic{equation}}
                \begin{eqnarray} }

\def\ie{\mbox{\it i.e.} }

\def\tr{{\rm tr}\,}

\def\Cbar{{\overline{C}}}

\def\Xbar{{\overline{X}}}

\def\Xt{{\tilde{X}}}

\def\khat{{\hat{k}}}

\begin{document}

\title{Masslessness of ghosts in
equivariantly gauge-fixed Yang--Mills theories}
\author{Maarten Golterman and Leah Zimmerman}
\affiliation{
Department of Physics and Astronomy,
San Francisco State University,
San Francisco, CA 94132, USA\\
{\tt maarten@stars.sfsu.edu},
{\tt lzimmer@stars.sfsu.edu}}

\begin{abstract}
We show that the one-loop ghost self-energy  in
an equivariantly gauge-fixed Yang--Mills theory vanishes at  zero momentum.  A ghost
mass is forbidden by equivariant BRST symmetry, and our calculation
confirms this explicitly.  The four-ghost self interaction which appears
in the equivariantly gauge-fixed Yang--Mills theory is needed in order
to obtain this result.
\end{abstract}

\pacs{11.15.Ha, 11.15.Bt}

\maketitle

\section{Introduction}

When one considers Yang--Mills (YM) theories in the continuum, one
needs to gauge fix the theory, and this is usually done
in the framework of BRST symmetry.  On the other hand, it is
well known that non-perturbatively gauge fixing is not needed if
the theory is defined on the lattice in terms of link variables
which are elements of a compact gauge group $G$.
The volume of a compact group is finite, and
if the theory is formulated on a finite lattice, it follows that
the path integral defining the (euclidean) YM theory is finite
without the need to fix a gauge.  However, the possibility of
gauge fixing a lattice YM theory non-perturbatively following the
standard BRST procedure would lead to interesting applications.
As an example, we mention the  construction non-abelian
chiral gauge theories on the lattice, where one recent approach
uses non-perturbative gauge fixing as a key ingredient to solving
the problem \cite{mgreview}.

The problem of defining a gauge-fixed lattice YM theory with
BRST symmetry was considered before \cite{hn1,hn2}.
In particular, it was shown that the lattice
partition function of a YM theory with
standard BRST symmetry vanishes identically on a finite lattice,
when formulated in terms of the link
variables \cite{hn2}.  This result extends trivially
to the (un-normalized) expectation value of any gauge-invariant
operator.  The result is a ``no-go" theorem saying that
YM theories cannot be gauge fixed on the lattice while maintaining
BRST symmetry.  Since the lattice is the only presently known
tool for defining YM theories non-perturbatively, this relegates
the standard BRST approach to a purely perturbative construct,
also on the lattice.

More recently, it has been shown that this problem
can be partially solved.  It was conjectured
that non-abelian theories with a gauge group $G$ can be
gauge fixed non-perturbatively down to a subgroup
$H\subset G$ provided that the Euler characteristic
of the coset manifold $G/H$ does not vanish \cite{schaden}.
Ref.~\cite{schaden}
worked this out in detail for $G=SU(2)$, $H=U(1)$, and it
was generalized in Ref.~\cite{gs} for $G=SU(N)$,  with
$H$ the maximal abelian subgroup $U(1)^{N-1}$.
(See Ref.~\cite{testa} for an alternative
approach dealing only with $G=U(1)$,
which is not the subject of this note.)

In this approach, ghosts are only introduced for coset
generators,
and the usual BRST transformation rules are changed to
maintain this constraint.  As a result,
the standard BRST algebra is changed into an
``equivariant" BRST (eBRST) algebra,  \ie a BRST algebra
in which the square of the BRST transformation is
a gauge transformation in $H$, instead of zero.
The gauge-fixing action is changed into an action invariant under eBRST
transformations.
The gauge-fixed theory is still $H$-gauge invariant, and
the modified BRST charge is still nilpotent on
$H$-gauge-invariant operators, which is sufficient to guarantee
unitary.  In fact,
correlation functions of gauge invariant
operators in the original un-fixed theory are identically
equal to those in the equivariantly gauge-fixed theory
\cite{gs}.

This alternative construction of a $G/H$ gauge-fixed lattice
YM theory leads to new interactions in the gauge-fixing
lagrangian.  Since $H$-gauge invariance is maintained,
one cannot choose  the Lorenz gauge, because all derivatives
need to be $H$-covariant.  A renormalizable gauge is thus
always non-linear, and leads to new gauge-field
self-interactions.  It turns out that the gauge-fixed
theory also contains four-ghost self-interactions, which play a
role in proving that the partition
function does not vanish in the equivariant case \cite{schaden,gs}.
No ghost mass term appears in the gauge-fixed action.
And indeed, one expects the ghosts to remain (perturbatively)
massless, if they are to play their usual role in guaranteeing
perturbative unitarity of scattering amplitudes of the massless gauge fields.

While the gauge-fixed theory is thus constructed to
have all the desired properties of a gauge theory, the way
this works out in detail is not trivial, because of the
presence of the new interaction terms introduced by the
equivariant gauge fixing.  In Ref.~\cite{gs} explicit
examples of one-loop unitarity of scattering amplitudes
were worked out.  These examples however do not involve the
ghost self-interaction.  In this note, we show by explicit
calculation that these ghost self-interactions are needed
to keep the ghost mass equal to zero to one loop.  In the standard
BRST case, this is rather trivial, due to the
fact that, in Lorenz gauge, a shift symmetry on the anti-ghost
prevents a ghost mass from being generated.  This shift
symmetry is not present in the eBRST case, and only eBRST
symmetry itself keeps the ghost mass at zero.  At one
loop, a
non-trivial cancellation between diagrams is required.  The one-loop
calculation thus serves as a non-trivial check on the validity
of the eBRST approach, and as an example
of the role of the ghost self-interaction.

While the main goal is to apply equivariant gauge fixing
to the construction of a non-perturbatively well-defined, gauge-fixed YM theory on the lattice,
it can also be formulated in the continuum.  Since here we will be concerned with
a perturbative investigation, we will first deal with the continuum case, and then
show that the same result is also obtained on the lattice.
We take $G=SU(N)$ and $H=U(1)^{N-1}$, the maximal abelian subgroup,
for which it was proven \cite{gs} that equivariant gauge fixing circumvents
the no-go theorem of Ref.~\cite{hn2}.
We devote Sec.~2  to the equivariantly gauge-fixed theory in
the continuum,  considering the ghost self-energy in
dimensional regularization (DR).  The fact that no ghost
mass is generated at one loop in DR results from the
cancellation of poles in $1/(d-2)$ between individual
diagrams.  In Sec.~3, we extend
our investigation to the lattice, showing that quadratic
divergences present in individual diagrams cancel, leading
to the same result.
The final section contains our conclusions.

\section{The continuum case}

Since only the coset $G/H$ will be fixed, we begin with splitting the gauge field
\begin{eqnarray}
\label{gaugefield}
V_\m&=&A_\m+W_\m\,,\\
A_\m&=&A^i_\m T^i\,,\ \ \ \ \ W_\m=W^\a_\m T^\a\,.\nonumber
\end{eqnarray}
The generators $T^i$ span the algebra $\ch$ for the subgroup $H$, while $T^\a$
span the coset space $\cg/\ch$, with $\cg$ the algebra for the group $G$.   The combined set $\{T^a\}=\{T^i,T^\a\}$ is normalized
by $\tr(T^a T^b)=\frac{1}{2}\d_{ab}$, and structure constants are defined by
$[T^a,T^b]=if_{abc}T^c$.   Indices $i,j,k,\dots$ ($\a,\b,\g,\dots$)
will be used to indicate generators in $\ch$ ($\cg/\ch$).
The field strength $F_{\m\n}$ is defined by $F_{\m\n}=\partial_\mu V_\n-\partial_\nu V_\m+i[V_\m,V_\n]$, as usual.

Since we leave $H$ un-fixed, the gauge condition $\cf(V)$ needs to be covariant under $H$,
and we choose
\begin{equation}
\label{gf}
\cf(V)=\cd_\mu(A)W_\mu\equiv\partial_\mu W_\mu+i[A_\mu,W_\mu]\,.
\end{equation}
For the same reason, we introduce only $\cg/\ch$-valued ghost and anti-ghost fields
\begin{equation}
\label{ghosts}
C=C^\alpha T^\alpha\,,\ \ \ \ \ \Cbar=\Cbar^\alpha T^\alpha\,.
\end{equation}
With these ingredients,
the continuum lagrangian for the equivariantly gauge-fixed YM theory is\footnote{This combines Eq.~(2.1) and a rewriting of
Eq.~(2.24) in Ref.~\cite{gs}.}
\begin{eqnarray}
\label{lgfonshell}
\cl&=&{1\over 2g^2}\tr(F_{\m\n}^2)+{1\over\xi g^2}\ \tr\left(\cd_\mu(A)W_\mu\right)^2\\
&&\hspace{-5mm}-2\,\tr\left(\Cbar\cd_\mu(A)\cd_\mu(A)C\right)
+2\,\tr\left([W_\mu,\Cbar]_\ch[W_\mu,C]_\ch\right)\nonumber\\
&&+i\,\tr\left((\cd_\mu(A)\Cbar)[W_\mu,C]+
[W_\mu,\Cbar](\cd_\mu(A)C)\right)\nonumber\\
&&+\x g^2\left(-{1\over 2}\tr(\Cbar^2C^2)+\tr(\Xbar X)-{1\over 4}\tr(\Xt^2)\right) \,,\nonumber
\end{eqnarray}
where $X$, $\Xbar$ and $\Xt$ are defined by
\begin{eqnarray}
\label{x}
X&\equiv&\left(iC^2\right)_\ch=2iT^j\ \tr(C^2T^j)\,,\\
\Xbar&\equiv&\left(i\Cbar^2\right)_\ch=2iT^j\ \tr(\Cbar^2T^j)\,,\nonumber\\
\Xt&\equiv&\left(i\{\Cbar,C\}\right)_\ch=2iT^j\ \tr(\{\Cbar,C\}T^j)\,.\nonumber
\end{eqnarray}
This on-shell lagrangian is invariant under the on-shell  eBRST
algebra.
For the construction of this lagrangian, and a discussion of eBRST
and other symmetries, we refer to Sec.~2 of Ref.~\cite{gs}.
Here we only note that a ghost mass
term is excluded by eBRST invariance, and not by any other symmetry of the gauge-fixed
theory.

We will now sketch the calculation of the ghost self-energy.  The relevant Feynman
rules for the vertices follow from (after rescaling $A^i_\m\to gA^i_\m$, $W^\a_\m\to gW^\a_\m$)
\begin{eqnarray}
\label{Feynman}
\langle A^i_\m(k)C^\a(p)\Cbar^\b(q)\rangle&=&-igf_{i\a\b}(p-q)_\m\,,\\
\langle W^\r_\m(k)C^\a(p)\Cbar^\b(q)\rangle&=&-{i\over 2}gf_{\r\a\b}(p-q)_\m\,,\nonumber\\
\langle A^i_\m(k)A^j_\n(l)C^\a(p)\Cbar^\b(q)\rangle&=&\nonumber\\
&&\hspace{-20mm}2\left[-{1\over 2}g^2
(f_{i\a\g}f_{j\b\g}+f_{j\a\g}f_{i\b\g})\d_{\m\n}\right]\,,\nonumber\\
\langle W^\r_\m(k)W^\s_\n(l)C^\a(p)\Cbar^\b(q)\rangle&=&\nonumber\\
&&\hspace{-20mm}2\left[{1\over 2}g^2
(f_{\r\a i}f_{\s\b i}+f_{\s\a i}f_{\r\b i})\d_{\m\n}\right]\,,\nonumber\\
\langle C^\a(p)\Cbar^\b(q)C^\r(k)\Cbar^\s(l)\rangle&=&
4\left[\x g^2\left({1\over 16}
f_{\a\r\g}f_{\b\s\g}\right.\right.\nonumber\\
&&\hspace{-30mm}\left.\left.
+{3\over 16}f_{\a\r i}f_{\b\s i}+{1\over 16}(f_{\a\b i}f_{\r\s i}
-f_{\a\s i}f_{\r\b i})\right)\right]\,,\nonumber
\end{eqnarray}
where on the left-hand side $\langle\dots\rangle$ denotes 1PI tree-level
correlation functions, and  the factors 2 and 4 outside the square brackets
on the right-hand side are combinatorial factors.
All momenta are in-going, and
delta functions for momentum conservation are understood.

The ghost and $W$ propagators are
\begin{eqnarray}
\label{props}
\langle C^\a(p)\Cbar^\b(q)\rangle&=&{\d_{\a\b}\over p^2}\,\d(p+q)\,,\\
\langle W^\a_\m(p)W^\b_\n(q)\rangle&=&
\d_{\a\b}\left({\d_{\m\n}\over p^2}+\left(\x-1\right)
{p_\m p_\n\over (p^2)^2}\right)\,\d(p+q)\,.\nonumber
\end{eqnarray}
In order to define the $A$ propagator, we have to gauge fix the
remaining abelian group $H$ as well.  We do this in
Lorenz gauge, adding a term ${1\over 2\a}(\partial_\m A^i_\m)^2$
to the lagrangian, leading to the propagator
\begin{equation}
\label{propA}
\langle A^i_\m(p)A^j_\n(q)\rangle=
\d_{ij}\left({\d_{\m\n}\over p^2}+\left(\a-1\right)
{p_\m p_\n\over (p^2)^2}\right)\,\d(p+q)\,.
\end{equation}
No other modifications are needed in order to gauge fix $H$;
in particular, no $\ch$-valued ghosts are needed.

The addition of ${1\over 2\a}(\partial_\m A^i_\m)^2$ to the
lagrangian breaks eBRST symmetry.  Reference~\cite{gs} discussed
how the Slavnov--Taylor identities are modified as a consequence,
and also how these identities can still be used to prove perturbative
unitarity of the theory after the two steps of $G/H$ and $H$ gauge fixing.
Here we are only interested in whether a non-vanishing ghost
mass is generated through radiative corrections, and
the relevant observations are rather simple.

On the lattice no gauge fixing of the abelian subgroup $H$ is
required, and eBRST symmetry does indeed preclude a non-zero ghost mass
term from appearing
in the $G/H$ gauge-fixed theory \cite{schaden,gs}.
The equivariantly gauge-fixed lattice
partition function is finite and
gauge invariant under $H$, and the ghost fields are
just a set of minimally-coupled ``matter" fields from the point of view
of $H$ symmetry.
Once the $H$ symmetry is also fixed
in order to set up perturbation theory, the pole mass of the ghost propagator
has to be $H$-gauge independent.  A proof of this can be given by adapting
a similar proof of the gauge independence of the perturbative pole mass of
quarks in QCD given in Ref.~\cite{ak}.  Combining these two facts leads to the
conclusion that the ghost mass vanishes to all orders in perturbation theory,
independent of the gauge parameter $\a$.
(For a very similar argument about the gauge independence of
S-matrix elements
in supersymmetric gauge theories in the Wess--Zumino
gauge, see Ref.~\cite{itoyamaetal}.)
Our explicit
calculation of the ghost self-energy at one loop will serve to
confirm these general observations.

At one loop, there are five diagrams contributing to the
ghost self-energy.  Two ``sunset"
diagrams involve the three-point vertices in
Eq.~(\ref{Feynman}), and have either an internal $A$ or $W$ line
in addition to an internal ghost line.
There are three tadpole diagrams involving the four-point vertices
in Eq.~(\ref{Feynman}), with an internal $A$, $W$ or ghost line.
The sunset diagram with an $A$ propagator is
\begin{eqnarray}
\label{ACCbar}
\Sigma_{AC\Cbar}(p)&=&g^2f_{\a\g i}f_{\b\g i}
\int {d^dk\over(2\pi)^d}{(2p-k)_\m(2p-k)_\n\over (k-p)^2}\nonumber\\
&&\times\ {1\over k^2}
\left(\d_{\m\n}+(\a-1){k_\m k_\n\over k^2}\right)\,.
\end{eqnarray}
The quadratic divergence present in this diagram, which in $d$ dimensions
shows up as a pole in $1/(d-2)$, can be made more
explicit by writing
\begin{eqnarray}
\label{ACCbarqd}
\Sigma_{AC\Cbar}^{\rm pole}(p)&=&\a g^2f_{\a\g i}f_{\b\g i}I(p)\,,\\
I(p)&=&
\int_0^1 dx \int {d^dk\over(2\pi)^d}
{2(1-x)(k^2)^2\over (k^2+\D)^3}\,,\nonumber
\end{eqnarray}
where $\D=x(1-x)p^2$.  The $A$ tadpole is
\begin{equation}
\label{Atadpole}
\Sigma_{A^2C\Cbar}(p)=-(d+\a-1)g^2 f_{\a\g i}f_{\b\g i}
\int {d^dk\over(2\pi)^d}\,{1\over k^2}\,.
\end{equation}
By inserting a factor $[k^2(k-p)^2]/[k^2(k-p)^2]$ into the integrand,
the $d=2$ pole part can be written as
\begin{equation}
\label{Atadpoleqd}
\Sigma_{A^2C\Cbar}^{\rm pole}(p)=-(d+\a-1)g^2 f_{\a\g i}f_{\b\g i}
\ I(p)\,,
\end{equation}
and we see that the $\a$-dependent part already cancels between
the two diagrams containing internal $A$ lines.

A very similar calculation leads to the following contribution to
the $d=2$ pole part from the two diagrams containing $W$ lines:
\begin{eqnarray}
\label{Wqd}
\Sigma_{WC\Cbar}^{\rm pole}(p)
+\Sigma_{W^2C\Cbar}^{\rm pole}(p)&=&\\
&&\hspace{-30mm}g^2
\left({1\over 4}\,\x\, f_{\a\g\r}f_{\b\g\r}
+(d+\x-1)\,f_{\a\g i}f_{\b\g i}\right)   I(p)\,.\nonumber
\end{eqnarray}
Finally, the contribution of the ghost tadpole is
\begin{equation}
\label{ghostqd}
\Sigma_{(C\Cbar)^2}^{\rm pole}(p)=-\x \left(
{1\over 4}f_{\a\g\r}f_{\b\g\r}+f_{\a\g i}f_{\b\g i}\right) g^2  I(p)\,.
\end{equation}
Clearly, all contributions to the pole near $d=2$ cancel, thus
proving that no ghost mass is generated at one loop.

\section{The lattice case}

In this section, we repeat the one-loop calculation of the previous
section, but we use the lattice instead of DR as a regulator.
There are two reasons for doing this:  First, equivariant gauge
fixing was designed to construct a lattice gauge-fixed gauge
theory with a form of BRST symmetry.  Second, since the lattice
regulator involves a physical cutoff $\propto 1/a$, with $a$ the
lattice spacing, the fact that no ghost mass is generated  follows
directly from the vanishing of the ghost self-energy
at zero momentum.  Unlike in DR, there is no need to go to
two dimensions to see the cancellation.
We will only highlight the differences between
the lattice and continuum calculations.

The ghost part of the lattice lagrangian is \cite{gs}
\begin{eqnarray}
\label{latlag}
\cl_{ghost}
&=&2\,\tr\left([T^i,D^+_\mu C_x][U_{x,\mu}T^iU^\dagger_{x,\mu},D^+_\mu\Cbar_x]\right)\\
&&\hspace{-10mm}-i\,\tr\left(2\{D^+_\mu C_x,\Cbar_x\}\cw_{x,\mu}
-\cw_{x,\mu}D^+_\mu\{\Cbar_x,C_x\}\right)\nonumber\\
&&\hspace{-10mm}+\x g^2\left(-{1\over 2}\tr(\Cbar^2C^2)
+\tr(\Xbar X)-{1\over 4}\tr(\Xt^2)\right) \,,\nonumber
\end{eqnarray}
in which
\begin{equation}
\label{covder}
D^+_\mu\Phi_x=U_{x,\mu}\Phi_{x+\mu}U^\dagger_{x,\mu}-\Phi_x\,.
\end{equation}
We note that this covariant difference
can be split into a (lattice) derivative
term $\partial^+_\m\Phi_x=\Phi_{x+\m}-\Phi_x$ and a term at least quadratic in the fields
if one expands $U_{x,\m}={\rm exp}(iV_{x,\m})$:
\begin{eqnarray}
\label{split}
D^+_\mu\Phi_x&=&\partial^+_\m\Phi_x+(U_{x,\mu}\Phi_{x+\mu}U^\dagger_{x,\mu}-\Phi_{x+\m})\\
&=&\partial^+_\m\Phi_x+i[V_{x,\m},\Phi_{x+\m}]+\dots\,.\nonumber
\end{eqnarray}

For our purpose, we only need three- and four-point vertices from this lagrangian,
since higher vertices do not contribute to the one-loop ghost self-energy.
Furthermore, since we are only interested in the ghost self-energy at zero
momentum, four-point vertices should not involve any (lattice) derivatives
of ghost or anti-ghost fields, and three-point vertices should involve at most
one derivative of a ghost or anti-ghost field.  This implies in
particular that we may replace $U_{x,\mu}T^iU^\dagger_{x,\mu}\to T^i$ in the
first term of Eq.~(\ref{latlag}), because  vertices coming from expanding the
link variables in this expression would either be three-point vertices involving
both derivatives on $C$ and on $\Cbar$, or four-point vertices involving at least one
such derivative.  It follows that the list of relevant vertices is similar to that in
Eq.~(\ref{Feynman}).  The ghost, $W$ and $A$ propagators are as given in Eqs.~(\ref{props},\ref{propA}),
but with $k_\m$ replaced by $\khat_\mu\equiv 2\sin{(k_\m/2)}$.

At zero momentum, it follows that one finds contributions
to the ghost self-energy in one-to-one correspondence to the continuum
contributions discussed in the previous section.   The results for the various
contributions are those given in Eqs.~(\ref{ACCbarqd},\ref{Atadpoleqd},\ref{Wqd},\ref{ghostqd}),
if one makes the replacements $d\to 4$, and
\begin{equation}
\label{latI}
I(p)\to {1\over a^2}\,\int_{-\pi}^{\pi}{d^4k\over(2\pi)^2}\,{1\over \khat^2}\,,
\end{equation}
where $\khat^2=4\sum_\mu\sin^2{(k_\mu/2)}$.  Adding up the contributions, we find that $\S(p=0)=0$, and
we conclude that also on the lattice no
ghost mass is generated at one loop, consistent with the DR calculation.
At non-zero momentum the calculation of the ghost self-energy is more
complicated than in the continuum, because for non-zero momentum
there are one-loop diagrams involving  ``lattice-artifact" vertices
(irrelevant vertices of order
$a^n$ with $n\ge 1$) which contribute to the ghost self-energy.

\section{Conclusion}

We have shown by explicit calculation that no ghost mass is generated in an
equivariantly gauge-fixed YM theory for the case that the subgroup $H$
is chosen to be abelian.
This is expected, because eBRST symmetry
forbids such a term in the lagrangian.  Nevertheless, the explicit calculation is
interesting for two reasons.  First, it provides an example of the role of the ghost self-interaction
which necessarily appears in the equivariantly gauge-fixed theory \cite{schaden,gs}.
Second, in order to investigate the equivariantly gauge-fixed theory
perturbatively, a further
gauge fixing of the subgroup $H$ is required.   Since this further gauge fixing
should not change particle masses in the equivariantly gauge-fixed theory,
one expects that masses are not affected by this additional
gauge fixing, even if it leads to a modification of the Slavnov--Taylor identities for
eBRST symmetry \cite{gs}.
This general argument is confirmed by our result.

We would like to comment on the extension of our results to higher orders in
perturbation theory.  In Sec.~2, we already gave a general argument as to why
one expects no ghost mass to be generated at any order in perturbation theory.
The explicit extension of the lattice perturbative calculation to higher loops
involves some technicalities.  The Lorenz gauge employed here to gauge
fix the abelian symmetry $H$ is non-linear on the lattice, because the
relation between $A_\m$ and $U_\m$ is non-linear.  Thus, even though
$H$ is abelian, the Lorenz gauge on the lattice requires another set of $\ch$-valued ghosts
in order to maintain BRST invariance for $H$ in lattice perturbation
theory.  (The no-go theorem of Ref.~\cite{hn2} is not relevant here,
just as it is not relevant when setting up standard perturbation
theory for YM theory.)
No such $H$-ghosts are required in
the continuum of course, so this is a pure lattice artifact.  The resulting new
vertices  on the lattice therefore correspond to irrelevant operators.  These new vertices
do not show up in the one-loop diagrams contributing to the self-energy of
the $\cg/\ch$-valued ghosts, which is why they did not appear in our
lattice calculation of the self-energy in Sec.~3.

\section*{Acknowledgements}

We thank Yigal Shamir for useful discussions and comments.
MG was supported in part by the US Department of Energy, 
and LZ was supported by the National Science Foundation funded GK-12 
fellowship program under grant number DGE-0337949.

\end{document}